\documentclass[sigconf]{acmart}

\newif\ifDEBUG

\DEBUGfalse

\newif\ifANONYMOUS

\ANONYMOUSfalse

\usepackage{soul}
\usepackage{xcolor}
\usepackage{multirow}
\usepackage{booktabs}
\usepackage{tabularx}
\usepackage{array}
\usepackage{array}
\usepackage{caption}
\usepackage{float}
\usepackage{xspace}
\usepackage{hyperref}
\usepackage{cleveref}
\usepackage{tcolorbox}
\usepackage{listings}

\newcommand{\hf}{Hugging Face\xspace}
\newcommand{\pypi}{PyPI\xspace}
\newcommand{\npm}{NPM\xspace}

\ifDEBUG
    \newcommand{\needcite}[1]{\hl{[#1]}}
    \newcommand{\needfig}[1]{\hl{[#1]}}
    \newcommand{\JJ}[1]
        {\textcolor{violet}{[JJ: #1]}}
    \newcommand{\WJ}[1]
        {\textcolor{blue}{[WJ: #1]}}
    \newcommand{\JD}[1]
        {\textcolor{purple}{[JD:#1]}}
    \newcommand{\AG}[1]
        {\textcolor{olive}{[AG:#1]}}
    \newcommand{\GKT}[1]
        {\textcolor{brown}{[GKT:#1]}}
    
    \newcommand{\NV}[1]
        {\textcolor{red}{[NV: #1]}}
    \newcommand{\CC}[1]
        {\textcolor{violet}{[CC: #1]}}
    \newcommand{\JY}[1]
        {\textcolor{cyan}{[JY: #1]}}
    \newcommand{\YT}[1]
        {\textcolor{red}{[YT: #1]}}

\else
    \newcommand{\needcite}[1]{}
    \newcommand{\needfig}[1]{}
    \newcommand{\JJ}[1]{}
    \newcommand{\WJ}[1]{}
    \newcommand{\GT}[1]{}
    \newcommand{\JD}[1]{}
    \newcommand{\AG}[1]{}
    \newcommand{\GKT}[1]{}
    \newcommand{\NS}[1]{}
    \newcommand{\NV}[1]{}
    \newcommand{\CC}[1]{}
    \newcommand{\JY}[1]{}
    \newcommand{\YT}[1]{}

\fi

\usepackage[compact]{titlesec}
    \usepackage{etoolbox}
    \makeatletter
    \patchcmd{\ttlh@hang}{\parindent\z@}{\parindent\z@\leavevmode}{}{}
    \patchcmd{\ttlh@hang}{\noindent}{}{}{}
    \makeatother

\newcommand{\myparagraph}[1]{\vspace{0.20cm}\noindent\textbf{#1} \noindent{}}

\usepackage{balance} 
\usepackage[font=small,labelfont=bf]{caption}
\usepackage{subcaption}
\makeatletter
\def\cl@chapter{}
\makeatother
\usepackage{cleveref}

\crefformat{section}{\S#2#1#3}
\crefname{figure}{Figure}{Figures}
\crefname{appendix}{Appendix}{Appendices}
\crefname{table}{Table}{Tables}
\crefname{algorithm}{Algorithm}{Algorithms}
\crefname{listing}{Listing}{Listings}
\crefname{theorem}{Theorem}{Theorems}
\crefname{thm}{Theorem}{Theorems}
\crefname{lemma}{Lemma}{Lemmata}
\crefname{equation}{Eqt.}{Eqts.}
\crefformat{Grammar}{Grammar #1}

\usepackage{enumitem}
\usepackage{booktabs}

\newcommand{\ie}{\textit{i.e.,} }
\newcommand{\eg}{\textit{e.g.,} }
\newcommand{\etal}{\textit{et al.}\xspace}

\newcommand{\code}[1]{{\small\texttt{#1}}\xspace}

\newcommand{\inlinequote}[1]{``\emph{#1}''}

\newcounter{finding}

\newcommand{\newfinding}{
  \refstepcounter{finding} 
  \textbf{Finding \thefinding.} 
}

\usepackage{caption}
\usepackage{subcaption}
\AtBeginDocument{
  \providecommand\BibTeX{{
    \normalfont B\kern-0.5em{\scshape i\kern-0.25em b}\kern-0.8em\TeX}}}

\copyrightyear{2024}
\acmYear{2024}
\setcopyright{rightsretained}
\acmConference[ESEM '24]{Proceedings of the 18th ACM / IEEE International Symposium on Empirical Software Engineering and Measurement}{October 24--25, 2024}{Barcelona, Spain}
\acmBooktitle{Proceedings of the 18th ACM / IEEE International Symposium on Empirical Software Engineering and Measurement (ESEM '24), October 24--25, 2024, Barcelona, Spain}
\acmDOI{10.1145/3674805.3686665}
\acmISBN{979-8-4007-1047-6/24/10}

\begin{document}

\title[What do we know about Hugging Face? A systematic literature review and quantitative validation of qualitative claims]{What do we know about Hugging Face? A systematic literature review and quantitative validation of qualitative claims}

\author{Jason Jones}

\email{jone2078@purdue.edu}
\orcid{0009-0005-7088-0597}

\affiliation{
  \institution{Purdue University}
  \streetaddress{610 Purdue Mall}
  \city{West Lafayette}
  \state{Indiana}
  \country{USA}
}

\author{Wenxin Jiang}

\email{jiang784@purdue.edu}
\orcid{0000-0003-2608-8576}

\affiliation{
  \institution{Purdue University}
  \streetaddress{610 Purdue Mall}
  \city{West Lafayette}
  \state{Indiana}
  \country{USA}
}

\author{Nicholas Synovic}

\email{nsynovic@luc.edu}
\orcid{0000-0003-0413-4594}

\affiliation{
  \institution{Loyola University Chicago}
  \city{Chicago}
  \state{Illinois}
  \country{USA}
}

\author{George K. Thiruvathukal}

\email{gthiruvathukal@luc.edu}
\orcid{0000-0002-0452-5571}

\affiliation{
  \institution{Loyola University Chicago}
  \city{Chicago}
  \state{Illinois}
  \country{USA}
}

\author{James C. Davis}

\email{davisjam@purdue.edu}
\orcid{0000-0003-2495-686X}

\affiliation{
  \institution{Purdue University}
  \streetaddress{610 Purdue Mall}
  \city{West Lafayette}
  \state{Indiana}
  \country{USA}
}

\newcommand{\structure}[1]{\ul{\textit{#1}}:}
\begin{abstract}  

\structure{Background}
Software Package Registries (SPRs) are an integral part of the software supply chain.
These collaborative platforms unite contributors, users, and code for streamlined package management.
Prior work has characterized the SPRs associated with traditional software, such as NPM (JavaScript) and PyPI (Python).
Pre-Trained Model (PTM) Registries are an emerging class of SPR of increasing importance, because they support the deep learning supply chain.
A growing body of empirical research has examined PTM registries from various angles, such as vulnerabilities, reuse processes, and evolution.
However, no synthesis provides a systematic understanding of current knowledge.
Furthermore, much of the existing research includes non-quantified qualitative observations.

\structure{Aims}
First, we aim to provide a systematic knowledge synthesis.
Second, we quantify qualitative claims.
 
\structure{Methods}
We conducted a systematic literature review (SLR).
We then observed that some of the claims are qualitative, lacking quantitative evidence.
We identify quantifiable metrics associated with those claims, and measure in order to substantiate these claims.

\structure{Results}
We identify 12 claims about PTM reuse on the HuggingFace platform, 4 of which lack quantitative support.
We tested 3 of these claims through a quantitative analysis, and directly compare the fourth with traditional software.
Our most notable findings are:
  (1) PTMs have a significantly higher turnover rate than traditional software, indicating more rapid evolution;
  and
  (2) There is a strong correlation between documentation quality and PTM popularity.

\structure{Conclusions}
Our findings validate several qualitative research claims with concrete metrics, confirming prior research. 
Our measures motivate further research on the dynamics of PTM reuse.

\end{abstract}





\maketitle

\section{Introduction}

\begin{figure*}[ht]
    \centering
    \includegraphics[width=0.90\textwidth]{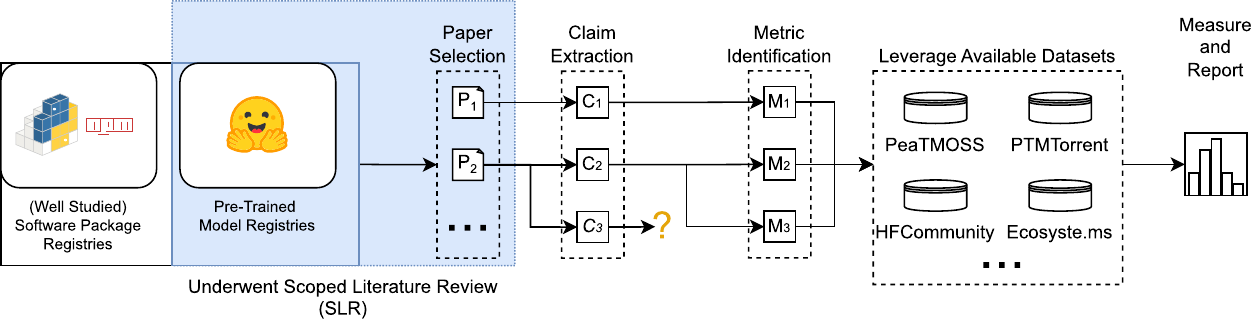}
    \caption{
      Overview of this work's context and approach.
      \ul{P}re-\ul{t}rained deep neural network \ul{m}odels (PTMs) are beginning to be reused, and accompanying empirical research is emerging.
      This work provides the first systematic literature review on PTM reuse.
      We extract the claims in prior work (RQ1) and provide quantitative evaluation of un-quantified and under-quantified claims (RQ2).
      \WJ{TODO: Add HF_Model_Metadata dataset to this figure.}
    }
    \label{fig:SLRQuantPipeline}
\end{figure*}

As the size and cost of developing deep learning (DL) models from scratch continue to rise, engineers are increasingly turning to adapt open-source Pre-trained Models (PTMs) as a cost-effective alternative~\cite{jiang_empirical_2022}. 
PTM registries facilitate the reuse of open-source models by providing packages that include pre-trained weights, configuration, and documentation~\cite{jiang_empirical_2023}.
Hugging Face has become a prominent PTM registry, comparable in popularity to traditional software registries like \npm and \pypi~\cite{jiang_empirical_2023}.
Understanding the characteristics of PTM registries such as Hugging Face is key to supporting effective and efficient software reuse in this emerging context.


Prior research has made significant strides in comparing PTM registries to traditional software package registries, addressing issues like carbon emissions, model selection, and vulnerabilities~\cite{jiang_empirical_2023, castano_analyzing_2023, kathikar_assessing_2023}.
However, there has been no systematic literature review that describes the current state of knowledge; such reviews advance the field by providing research agendas.
Our study thus contributes in three ways.
First, we provide the first systematic review of knowledge on PTM registries.
Second, we propose quantitative metrics for existing qualitative insights, enabling a more robust validation of existing claims about PTM registries.
Lastly, we validate or challenge previous qualitative insights through quantitative analysis.

As illustrated in~\cref{fig:SLRQuantPipeline},
our method has two parts.
First, we conduct a systematic literature review (SLR) to extract existing knowledge (claims) about Hugging Face.
Second, we identify un-quantified and under-quantified claims, and provide metrics and measurements using existing datasets. 

Our SLR extracted 12 distinct claims about HuggingFace, of which 4 lacked large-scale quantitative evidence. 
After defining metrics, we supported 2 of these with a large-scale measure; and found ambivalent evidence for a third.
The fourth was too imprecise to operationalize.
To summarize our findings:
  (1) Our measurement of library usage indicated a preference for using the \code{Transformers} library when creating descendents of models, with 80\% of descendents of models choosing to utilize \code{Transformers} ;
  (2) HuggingFace has a significantly higher turnover rate than traditional package registries;
  (3) Model popularity correlates with descendent count, supporting claims that popularity drives model selection;
  and
  (4) Documentation quality correlates with PTM popularity. 

\ul{Our contributions are:}
\vspace{-0.05cm}

\begin{itemize}[leftmargin=0.37cm]
    \item We conduct a systematic literature review on PTM reuse in Hugging Face, summarizing the claims of prior work (\cref{sec:RQ1}).
    \item We operationalize qualitative claims with quantitative measurements, and then validate most prior qualitative findings, via comparison of our quantitative measurements of \hf to representative traditional SPRs (\cref{sec:RQ2}).
    \item We recommend future work on analyzing and maintaining datasets for furrther study of the PTM supply chain (\cref{sec:discussion}).
\end{itemize}

\vspace{-0.05cm}
\noindent
\ul{\textbf{Significance:}}
Prior work has made both quantitative and qualitative claims about software engineers' reuse of PTMs.
We systematically synthesized this knowledge through a literature review, and developed quantifiable metrics to corroborate qualitative claims.
Our findings confirm several qualitative results about PTM reuse, and also quantify the dynamic reuse environment within the PTM ecosystem.
Our results will inform research infrastructure, and new metrics to guide and refine PTM development and reuse.

{
\renewcommand{\arraystretch}{0.9}
\begin{table*}
\centering
\caption{
Example metrics used to characterize traditional software package registries.
The metrics in this table are used as guidance when developing metrics to measure the claims in \cref{table: SLR-claims}.
}
\small
\begin{tabular}{m{0.15\linewidth} m{0.3\linewidth} m{0.27\linewidth} m{0.2\linewidth}} 
\toprule
\textbf{Metric} & \textbf{Description \& Implication} & \textbf{Package Registries} & \textbf{Example Works}  \\ 
\midrule
User Reach           &  The amount of ecosystem controlled by a fraction of its maintainers.
  & NPM, \pypi, Cargo, Elm, CRAN                                      & \cite{ahmed_case_2017, bommarito_empirical_2019, liu_demystifying_2022, blanthorn_evolution_2019, zhang2020companies, gu2023investigating,  zimmermann2019small} \\ 

\addlinespace
Dependency Degree    & The number of dependencies among software packages. & NPM, PyPi, Cargo, CPAN, CRAN, NuGet, Packageist, RubyGems, Maven & \cite{zerouali2019formal, pinckney2023large, wittern_look_2016, zerouali2018empirical, bommarito_empirical_2019, decan2019empirical, liu_demystifying_2022, blanthorn_evolution_2019, gu2023investigating, zerouali_diversity_2019, decan2018evolution, zimmermann2019small, stringer2020technical, wu_understanding_2023, kula2014visualizing} \\ 

\addlinespace
Popularity           & The frequency of use for top packages, indicating the concentration of usage.           & NPM, \pypi, OpenStack                                      & \cite{wittern_look_2016, dey_are_2018, zerouali_diversity_2019} \\ 
\addlinespace
Technical Lag        & The delay in adopting new updates.          
& NPM, \pypi, Maven, Cargo                                          & \cite{zerouali2019formal, pinckney2023large, wittern_look_2016, zerouali2018empirical, decan2018evolution, stringer2020technical} \\ 
\bottomrule
\end{tabular}
\label{table:metrics}

\end{table*}
}

\section{Background and Related Work}

We discuss how software package registries facilitate the reuse of software, and deep neural network-specific considerations for package registries that contain Pre-Trained Models (\cref{sec:background-SPR}). 
We describe how software package registries have been measured in previous works, and the limited quantification of PTM registries (\cref{sec:background-SPRMeasurements}).

\subsection{Software Package Registries, Old and New}
\label{sec:background-SPR}
\begin{figure}
    \centering
    \includegraphics[width=0.95\columnwidth]{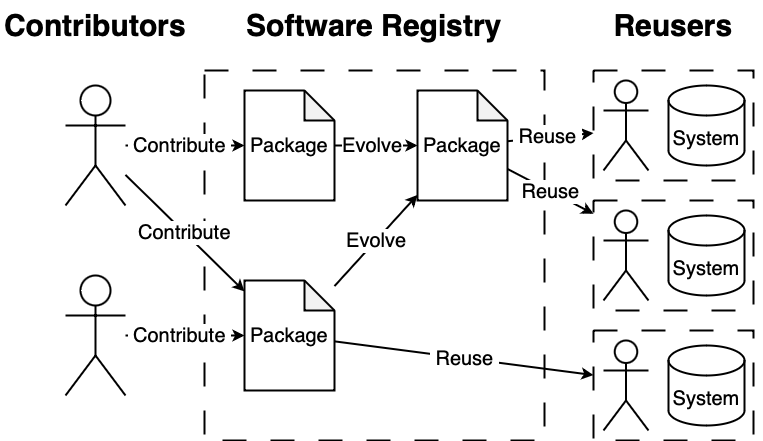}
    \caption[The Software Supply Chain]{
    The software supply chain.
    Software package registries connect package authors to reusers, accelerating system development.
    }
    \label{fig:ssc}
\end{figure}

\subsubsection{General Concepts.}


Software Package Registries (SPRs) serve as collaborative hubs that connect package contributors to package reusers, facilitating software reuse. 
\cref{fig:ssc} depicts how contributors collaborate on packages that are reused as components further down the software supply chain. 
For package contributors (maintainers), SPRs provide tools for package lifecycle management:
  bundling,
  versioning,
  hosting,
  and so on~\cite{npm, pypi}.
For package reusers, SPRs assist in package search and retrieval, and often provide direct or indirect measures of package quality through user engagement tools like comments, likes, and ratings. 
SPRs typically provide both existing components, and metadata about their provenance, performance, and other attributes that constrain the potential for reuse.


\subsubsection{SPRs for Deep Neural Networks.}

Deep Neural Networks (DNNs) provide state-of-the-art performance for many tasks, such as image recognition in autonomous vehicles~\cite{garcia2020AVbugstudy} and AI voice assistant systems~\cite{nasirian2017ai}. 
Developing and training these models from scratch requires significant time and resources~\cite{rae2021scaling, hoffmann2022training, patterson2022GoogleMLTrainingCarbonFootprint}.
For example, a \code{Llama-2-70B} model needs 1720K GPU hours to train from scratch~\cite{touvron2023llama2}.
Engineers thus often use Pre-Trained Models (PTMs). 
They can then undertake a much shorter training period to fine-tune the models for specific tasks~\cite{jiang_empirical_2023, davis2023reusing, Jiang2024Challenges}.

Since PTMs are reused components, PTM package registries (\eg Hugging Face~\cite{wolf_huggingfaces_2020}, PyTorch Hub~\cite{pytorchhub}, ONNX Model Zoo~\cite{onnxmodelzoo}) have emerged to support the development and reuse of trained DNNs for use in the development of AI systems~\cite{davis2023reusing, jiang_empirical_2022}.
PTM package registries are modeled on traditional SPRs, with some extensions for DNN-specific properties.
PTM packages include traditional package components such as documentation and dependencies.
However, they also include additional DNN-specific components, such as pre-trained weights, training dataset(s), and model architecture~\cite{jiang_empirical_2023}.

\begin{figure}[ht]
    \centering
    \includegraphics[width=0.95\columnwidth]{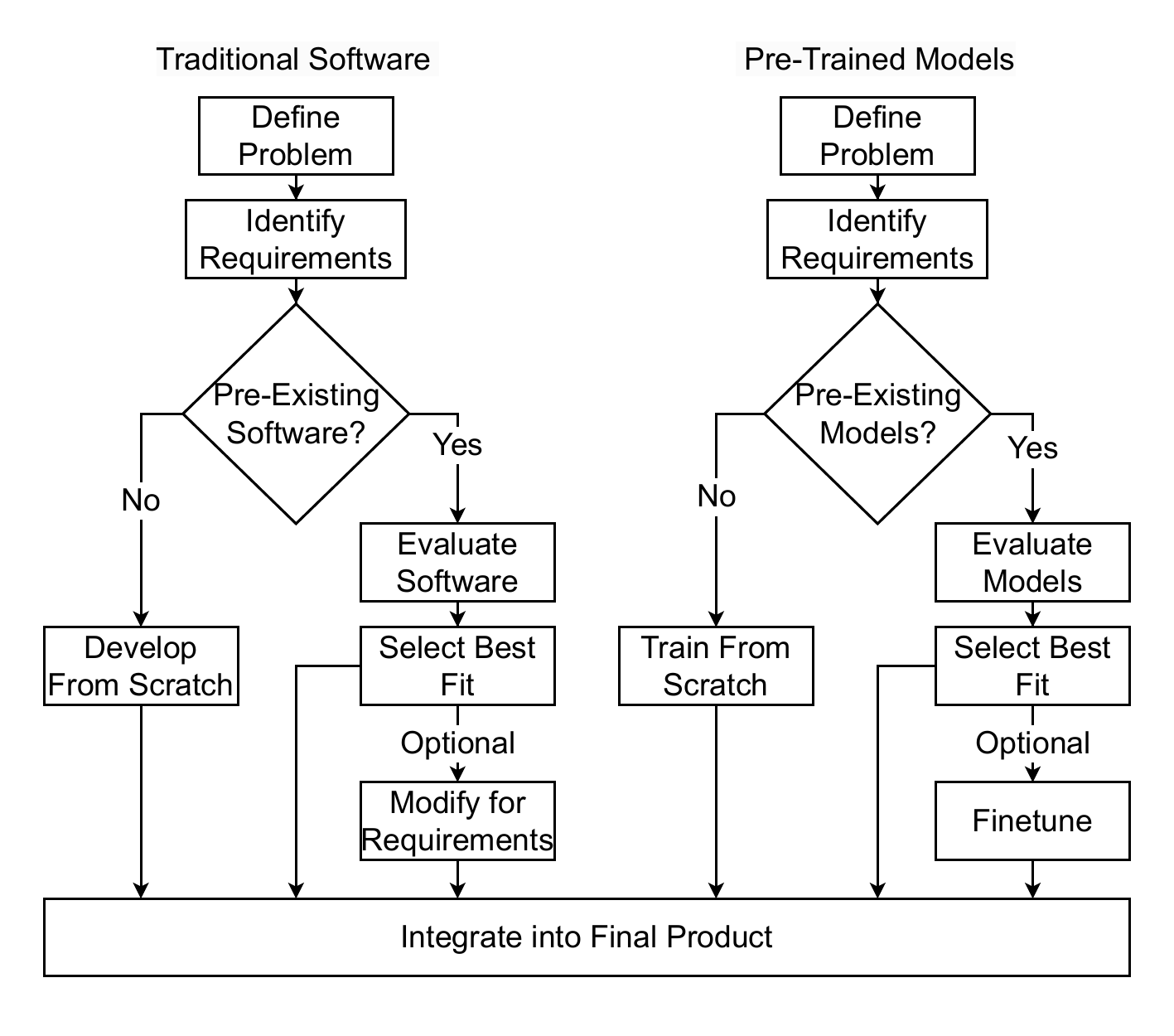}
    \caption{
    Reuse processes of traditional software and of PTMs, as reported by Jiang \etal~\cite{jiang_empirical_2023}. 
    Reuse processes are similar, suggesting that SPR measurements and trends may be similar.
    \WJ{Taylor: Missing important evaluation loop here which is the major different. Otherwise these two reuse processes are mostly the same}
    \JD{Agree, let's add the loop.}
    }
    \label{fig:ReuseComparason}
\end{figure}

Within the context of PTMs, the term ``software reuse'' refers to the reuse of models along with their training configurations and weights, \ie reuse of the PTM packages. 
\cref{fig:ReuseComparason} compares the reuse process between traditional packages and PTM packages.
Reusing DNNs as PTM packages mirrors traditional practices of software package reuse, where existing software components are integrated and adapted rather than built from scratch~\cite{jiang_empirical_2022, gong_what_2023}.
This practice can enhance efficiency and reducing development time~\cite{gopalakrishna_if_2022}. 
This adaptation involves not only the models but also their pre-trained states, which can be adjusted for specific applications, thereby constituting a combination of model reuse and customization rather than traditional software package reuse alone.

\subsection{Measuring Software Package Registries}
\label{sec:background-SPRMeasurements}

Research on traditional software package registries has typically sought to quantify various aspects of SPRs~\cite{zerouali2019formal, decan2018evolution, german_understanding_2010}.
In contrast, PTM registry research has often focused on qualitative and small-scale quantitative measurements to date.
This difference is appropriate given the need to provide qualitative descriptions of phenomena prior to measurement of them.
However, this difference also highlights the novel aspects of PTM registries and emphasizes the need for the systematization of existing knowledge.

\subsubsection{General Comparisons.}
The measurement of software package registries is an established research area within traditional software. 
\cref{table:metrics} shows the metrics used in prior work \eg user reach~\cite{zimmermann2019small}, license~\cite{german_understanding_2010}, and technical lag~\cite{decan2018evolution}. 
These studies provide critical insights for software engineers, aiding in effective package selection and reuse, and for researchers, deepening understanding of software supply chains and registry dynamics.

This research approach has been adapted to PTM registries recently~\cite{jiang_naming_2024, jiang_empirical_2023}. 
Early studies have explored how PTMs are reused and adapted, examining both qualitative and quantitative aspects. 
For instance, research has started to analyze contribution patterns and the reuse dynamics specific to PTM registries, such as those found in TensorFlow Hub and Hugging Face \cite{jiang_empirical_2023, miao2017modelhub, davis2023reusing}.
However, there are still gaps in understanding the evolution and reuse patterns in the PTM registries~\cite{jiang_empirical_2023, jiang_peatmoss_2024}. 
Our work bridges this gap by providing additional quantitative measurements and comparing our results to traditional SPRs.

\subsubsection{Explaining Phenomena Before Quantifying Them.}
Despite these efforts, there remains a gap in research that connects qualitative observations with quantitative measurements in PTM registries. 
Recent studies within traditional SPRs typically start with qualitative observations that are later supported by quantitative data. 
Issues such as package obsolescence and the spread of vulnerabilities, initially observed qualitatively, have been rigorously quantified in ecosystems such as NPM~\cite{mujahid2023characteristics, abdalkareem2017developers}. 
These investigations provide a basis for understanding how aspects of software registries impact the practices of software development.

In the research domain of PTM registries, similar investigations are needed. 
For example, Jiang \etal provide a comprehensive analysis on the risks while using PTMs, and the PTM reuse process~\cite{jiang_empirical_2022, jiang_empirical_2023}. 
As a follow-up work, they also collected both qualitative and quantitative data on PTM naming practices~\cite{jiang_naming_2024}. 
This line of research is crucial for understanding PTM package registries.
However, their qualitative insights have not been substantiated by quantitative analysis.
We address this gap by conducting a systematic literature review and deriving quantifiable metrics to assess the claims from previous studies.

\section{Knowledge Gap and Research Questions}
\label{sec:KnowledgeGapandRQ}

To summarize the knowledge gap, we lack a cohesive understanding on PTM package reuse that synthesizes prior research on Hugging Face.
Additionally, there is a lack of some quantitative measurements, which reduce our understanding of the registry.
To address the gap of knowledge synthesis, we ask:

\begin{itemize}[leftmargin=26pt, rightmargin=5pt]
\item[\textbf{RQ1}] What claims about package reuse on Hugging Face are made by prior research?

\end{itemize}

By gathering and analyzing these claims, we convert qualitative data into a set of quantifiable metrics. 
This effort enriches our understanding of Hugging Face as a PTM platform and enhances the comparability of data across different software ecosystems. 
Specifically, future researchers are enabled to establish metrics that facilitate the comparison of Hugging Face and traditional software registries similarities and differences in package reuse practices.

Answering RQ1 also prepares us for further empirical inquiry:

\begin{itemize}[leftmargin=26pt, rightmargin=5pt]
\item[\textbf{RQ2}] 
Do the qualitative claims about package reuse (PTM) on Hugging Face hold up when quantified?

\end{itemize}

\subsection{Overview of Methodology}

\cref{fig:SLRQuantPipeline} shows the overview of our work's context and approach.

\begin{enumerate}[leftmargin=1.5em]
  \item We conducted a systematic literature review on PTM reuse, compiling a comprehensive list of claims about package reuse.
  \item We synthesized these claims into quantifiable metrics, took measurements, and compared them to previous findings.
  \item We compared measurements of PTM packages to those of traditional packages to see differences in reuse patterns.
\end{enumerate}

The detailed method per RQ is in the corresponding sections.

\section{RQ1: Prior Package Reuse Claims}

\begin{tcolorbox} [width=\linewidth, colback=yellow!30!white, top=1pt, bottom=1pt, left=2pt, right=2pt]

  \newfinding The Systematic Literature Review (SLR) identified 12 quantifiable claims about package reuse on Hugging Face.

  \newfinding The claims are distributed among five quantifiable categories of methods: small- and large-scale quantitative measurements, qualitative surveys, interviews, and case studies.

\newfinding After this classification, four quantitatively unevaluated claims remained, within the categories of
    \emph{design trends}, 
    \emph{documentation and understanding}, and
    \emph{selection considerations}. 
  This finding shows a gap that we address in RQ2.

\end{tcolorbox}

\label{sec:RQ1}


To enhance our understanding of the existing claims related to PTM reuse within package registries, we conducted a systematic literature review (SLR) following established empirical standards~\cite{ralph2020empiricalStandards, ralph2022pavingtheWay4MatureSecondaryResearch, borenstien2009introduction, keele_guidelines_2007}.
The SLR protocol was designed to ensure a comprehensive and unbiased review of the literature, with a focus on identifying quantifiable metrics that could be used to assess claims about PTM reuse on Hugging Face.

The need for quantifiable metrics in this study stems from the importance of objectively measuring and comparing aspects of PTM reuse across different studies and contexts.
These metrics allow us to
  validate qualitative observations with quantitative data,
  identify trends and patterns in PTM reuse practices,
  and
  compare findings across different studies and ecosystems.

\begin{figure}[h]
    \centering
    \includegraphics[width=0.98\columnwidth]{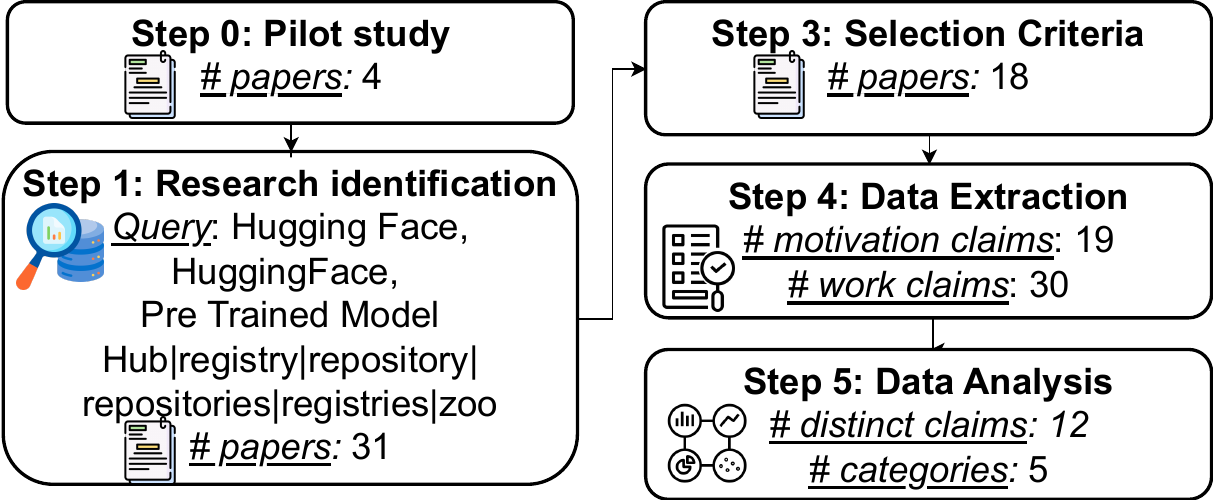}
    \caption{
    Systematic literature review process and results.
    }
    \label{fig:SLRProcess}
\end{figure}

Our initial step involved a pilot study to define the scope of our review (\cref{sec:SLR-PilotStudy}).
A systematic literature review entails five steps: 
    identification of research, 
    study selection, 
    study quality assessment, 
    data extraction, and 
    data analysis~\cite{keele_guidelines_2007, moher2010preferredSLR}.
\cref{fig:SLRProcess} illustrates the results of each step.
We detail them next.

\subsection{Methods}

\subsubsection{Pilot Study.}
\label{sec:SLR-PilotStudy}

We define the scope of our review by conducting a pilot study.
In the pilot study, we first search for papers about ``pre-trained model reuse'' using Google Scholar and looked at the first three results, which were~\cite{jiang_empirical_2022, jiang_empirical_2023, gong_what_2023}. 
The papers indicated that Hugging Face is the only ``open'' model registry~\cite{jiang_empirical_2022} and is the most popular model registry. 
Additionally, Hugging Face hosts the largest number of PTM packages, and provides useful tools to facilitate PTM reuse. 
We then decided to scope down our study on Hugging Face model registry specifically to represent the PTM supply chain, as indicated by Jiang \etal~\cite{jiang_peatmoss_2024, jiang_empirical_2023}.

\subsubsection{Search Strategy and Query.}
\label{sec:SLR-SearchQueries}

The goal of our search is to identify papers that are relevant to the categories of 
    PTM reuse,
    the PTM supply chain, or 
    the PTM ecosystem.

Informed by our pilot study (~\cref{sec:SLR-PilotStudy}), the final search query we used is indicated in~\cref{fig:SLRProcess}.
These search queries gave us 45 papers.
We then removed duplicate entries, reducing the number of papers to 31.
To verify the efficacy of our search queries, we employed papers from the pilot study as benchmarks to assess each query's retrieval effectiveness. 
We ensured that all papers identified in the pilot study were also retrieved by our final search query.
This step was essential to confirm the robustness of our search strategy, ensuring it was capable of capturing the most relevant studies. 
Such a comprehensive approach allowed for an exhaustive review of the literature concerning PTM reuse within its operational ecosystem.

\subsubsection{Selection Criteria.}
\label{sec:SLR-SelectionCriteria}





The goal of our selection criteria is to identify the most relevant and rigorously supported research that specifically addresses the context and impact of PTM reuse.
We developed and applied a set of clearly articulated criteria to ensure consistency and reproducibility in our selection process.

\myparagraph{Inclusion Criteria}
We applied the following inclusion criteria to our systematic literature review:
The paper must describe the reuse of models within a specific PTM registry, particularly Hugging Face.
Additionally, the study should focus on PTM reuse practices, challenges, or impacts.
To ensure academic rigor, we only included papers published in peer-reviewed journals or conference proceedings.
Lastly, to maintain relevance to current practices, we limited our search to publications within the last five years (2019-2024).

\myparagraph{Exclusion Criteria}
We applied the following exclusion criteria in our systematic literature review:
Papers that only apply PTMs to specific tasks without discussing reuse practices were excluded.
We also omitted non-primary sources, such as literature reviews or opinion pieces, to focus on original research.
Additionally, we excluded studies whose claims were not substantiated directly through qualitative or quantitative methods, ensuring that our review was based on well-supported findings.

Application of these criteria reduced our initial set of 31 papers to 18 high-quality, relevant studies for in-depth analysis.

{
\renewcommand{\arraystretch}{0.9}
\begin{table*}[ht!]
    \centering
    \caption{
    The detailed definition and examples of each claim category we extracted from the literature review.
    }
\small
    \begin{tabular}{lp{0.32\textwidth}p{0.49\textwidth}}
    \toprule
         \textbf{Claim Category} & \textbf{Definition} & \textbf{Examples}\\
     \midrule
          Motivation claims & Articulates the rationale behind a paper's problem statement, illustrating why the work is significant and worthy of investigation. & 
         \inlinequote{The reuse of pre-trained models introduces large costs and additional problems of checking whether arbitrary pre-trained models are suitable for the task-specific reuse or not.}~\cite{gong_what_2023} \\
         
         \addlinespace
          & & \inlinequote{With the commodification of AI, and NLP in particular...[we] need easy to use, no-code tools for understanding AI artifacts.}~\cite{piktus2023gaia} \\

\midrule
         Work claims & Refers to the assertions a paper makes based on its collected data and analyses. & \inlinequote{Hugging Face's popularity has exponentially increased over time, which is evident from the upward trends in the number of new models, likes, commits, unique authors, and discussions aggregated monthly.}~\cite{castano_analyzing_2023} \\

\addlinespace
         &  &  \inlinequote{Engineers follow specific naming practices and encounter challenges that are specific to PTM naming.}~\cite{jiang_naming_2024} \\
         \bottomrule
    \end{tabular}
    \label{tab:ClaimCategories}
\end{table*}
}

\subsubsection{Data Extraction.} 

\label{sec:SLR-DataExtraction}

Once we identified the most relevant and rigorously supported research that specifically addresses the context and impact of PTM reuse, the next step is to extract ``claims'' from these papers that provide evidence of qualitative or quantitative methods that could inform our quantitative study.
A ``claim'' in this context refers to a statement or assertion made in a research paper, which is supported by evidence.

The data extraction involve four steps.
\emph{First}, two co-authors went through a paper from the pilot study together to get an agreement of the data extraction process. 
The goal of this process was to identify and extract all claims that might be tangentially related to PTM reuse.
\emph{Second}, they individually extracted the claims from the papers.
This involved reading each paper to identify the key claims, with an emphasis placed on the abstract, introduction, and, if available, finding boxes of each paper.
Exact quotations from the papers were extracted.
\emph{Third}, the two co-authors met and presented their extracted claims from each paper, with an explanation of why it was chosen, and a discussion of the relevance of the claim if it was not immediately obvious.
A total of 256 claims were discussed in this step.
\emph{Finally}, for each paper, they discussed which claims were the most descriptive of \textit{PTM reuse} and discarded the rest.
This selection step resulted in a total of 49 claims.

\subsection{Analysis and Results}
\label{sec:SLR-Results}


We categorized these claims into two categories:
(1) ``\emph{Motivation claims}'' and (2) ``\emph{Work claims}''. 
The detailed definitions and examples of each claim are shown in \cref{tab:ClaimCategories}.
This classification process yielded 19 \emph{motivation claims} and 30 \emph{work claims}. 
One of our goals (RQ2) was to substantiate prior findings with further quantitative measurements, so we chose to focus more deeply on the work claims. 
Within this subset, we identified overlapping claims and consolidated them, resulting in a refined set of 12 distinct claims. 
These consolidated claims are detailed in \cref{table: SLR-claims}.

The primary aim of our SLR is to summarize the existing claims about package reuse on Hugging Face and to extract quantifiable measurements from these claims.
From the consolidated set of 12 claims, we categorized the basis of each claim into one of five methods: small- and large-scale quantitative measurements, qualitative surveys, interviews, and case studies.
\emph{Small-scale measurements} involved less than 10\% of the population, while \emph{large-scale measurements} covered more than 10\%.
After this classification, five quantitatively unevaluated claims remained: one concerning \emph{design trends}, two regarding \emph{selection considerations}, and two about \emph{documentation and understanding}.
The categorization result and extracted themes are detailed in \cref{table: SLR-claims}. 

Our comprehensive knowledge synthesis addresses RQ2 (\cref{sec:RQ2}), revealing several key insights into PTM reuse research.
We observe a significant shift towards large-scale, data-driven analysis, complemented by diverse research methods that collectively strengthen the field's knowledge base. 
The consolidation of 30 initial claims into 12 distinct work claims indicates both a convergence of findings and a need for validation through replication studies.
Notably, we identified five quantitatively unevaluated claims, highlighting critical areas for future empirical investigation, particularly in design trends, selection considerations, and documentation practices.

Our findings have significant implications for PTM reuse research and practice. For researchers, we provide a roadmap highlighting areas requiring rigorous quantitative evaluation.
Practitioners can leverage our identified themes to inform PTM selection and implementation decisions.
The emphasis on documentation-related claims underscores its critical importance in the PTM ecosystem for both groups.
This comprehensive analysis lays the foundation for targeted, impactful research, contributing to the maturation of the rapidly evolving PTM reuse field.


{
\renewcommand{\arraystretch}{0.9}
\begin{table*}[ht]
    \centering
    \caption{
    Consolidated themes and claims collected from our systematic literature review. Small-Scale measurements refer to measurements made on a selection of a population less than 10\% of the overall size, Large-scale measurements are measurements made on a selection of a population that is more than 10\%, Survey refers to a survey, Interviews refer to interviews, Case study refers to either an examination of a specific case or the creation of a model to verify the claim.
    $\dagger$ : The claim basis is not a large-scale quantification.
    }
\small
    \begin{tabular}{m{0.15\linewidth} m{0.47\linewidth} m{0.2\linewidth} m{0.08\linewidth}}
    \toprule
       \textbf{Themes}  &  \textbf{Claims} & \textbf{Claim Basis} & \textbf{Works} \\
    \midrule

Trends in Design & A small group of contributors owns popular models. & Large-Scale Measurement & \cite{castano_analyzing_2023} \\
     \addlinespace
     & The Transformers library increases the accessibility of PTM creation and downstream reuse. & Large-Scale Measurement, Case Study & \cite{kathikar_assessing_2023, wolf_huggingfaces_2020, jiang_peatmoss_2024, castano_analyzing_2023} \\
     \addlinespace
     & $\dagger$ \textit{(1) The Transformers library improves the process of PTM evolution}. & \textbf{Case Study} & \cite{wolf_huggingfaces_2020} \\
     \addlinespace
     & Forking repositories introduces low severity vulnerabilities. & Large-Scale Measurement & \cite{kathikar_assessing_2023} \\
     \addlinespace
    \midrule
    Selection Considerations & $\dagger$ \textit{(4) DL-specific attributes such as model architecture, performance, reproducibility, and portability affect PTM selection and reuse.} & \textbf{Interviews} & \cite{jiang_empirical_2023} \\
    \addlinespace
     & $\dagger$ \textit{(2) The traditional attribute of Popularity affects PTM selection and reuse more than Maintenance and Quality.} & \textbf{Interviews} & \cite{jiang_empirical_2023} \\
     \addlinespace
    \midrule
    Repository Lifecycle and Maintenance &  Models receive perfective maintenance over time, with high-maintenance models tending to be more popular, larger, and better documented. & Large-Scale Measurement & \cite{castano_analyzing_2023} \\
    \addlinespace
    \midrule
    Documentation and Understanding & Model properties are under-documented across Hugging Face. & Small-Scale and Large-Scale Measurement, Survey & \cite{kathikar_assessing_2023, taraghi_deep_2024, montes_discrepancies_2023, gong_what_2023} \\
    \addlinespace
    & $\dagger$ \textit{(3) Documentation quality impacts model selection}. & \textbf{Survey, Case Study} & \cite{taraghi_deep_2024, montes_discrepancies_2023, castano_lessons_2024} \\
    \addlinespace
    & Naming models is inconsistent and can inadequately represent model architectures. & Large-Scale Measurement, Survey & \cite{jiang_naming_2024} \\
    \addlinespace
    & Dataset documentation is correlated with dataset popularity. & Small-Scale Measurement & \cite{yang_navigating_2024} \\
    \addlinespace
    \midrule
    Downstream Usage &  Hugging Face is an exponentially growing platform. & Large-Scale Measurement & \cite{castano_analyzing_2023} \\
    \addlinespace
    \bottomrule
    \end{tabular}
    \label{table: SLR-claims}
    
\end{table*}
}

\section{RQ2: Qualitative Claim Validation}
\label{sec:RQ2}

\begin{tcolorbox} [width=\linewidth, colback=yellow!30!white, top=1pt, bottom=1pt, left=2pt, right=2pt]
    \newfinding Our study shows that the Transformers library is preferred in over 80\% of PTM descendants, surpassing PyTorch. The rise of the SafeTensors library underscores a shift toward prioritizing security in PTM development.

    \newfinding \cref{fig:turnover} reveals that Hugging Face has a significantly higher package turnover rate than traditional software registries indicative of a fast-paced, innovation-driven PTM ecosystem.

\newfinding Model popularity correlates with descendant count, indicating that while popular models have more descendants, other factors influence model selection decisions.

\newfinding  There is a strong correlation between documentation quality and PTM popularity, with the top 1000 models significantly outperforming the bottom 1000 in documentation, highlighting its importance in model selection.

\end{tcolorbox}



To answer this question, we first derive quantifiable metrics from the claims we extracted from our SLR (\cref{sec:RQ1}) on Hugging Face (\cref{sec:RQ2-DerivedMetrics}).
Then we present the available datasets and the specific data we used from each (\cref{sec: RQ2-Dataset}).
Subsequently, we present our methods and results for measurement on metrics for each claims (\cref{sec:RQ2-Claim1}--\cref{sec:RQ2-Claim3}).

\subsection{Metrics Developed from Claims}
\label{sec:RQ2-DerivedMetrics}
In this section, we develop quantifiable metrics from the claims identified in our SLR (\cref{sec:RQ1}).
These metrics are designed to test hypothesized results inferred from the claims, drawing on established software engineering practices and previously used metrics where applicable.
We prioritized widely recognized, frequently cited metrics that have been implemented across various traditional software registries.
This approach grounds our metrics in established methodologies, ensuring robust and meaningful insights.
\cref{tab:Claim-Metrics} shows the relationship between extracted claims, corresponding metrics, and expected measurements, validating the claims and contextualizing their implications for package reuse on \hf.

\subsection{Available Datasets}
\label{sec: RQ2-Dataset}

The section presents the PTM package datasets (\cref{sec:Dataset-PTM}) and traditional software package dataset (\cref{sec:Dataset-TradSW}) we used in our work.

\subsubsection{PTM Datasets.}
\label{sec:Dataset-PTM}

In the PTM literature, there are four datasets available publicly: HF Model Metadata~\cite{vanstrien_hf_model_metadata}, PTMTorrent~\cite{jiang_ptmtorrent_2023}, HFCommunity~\cite{ait2023hfcommunity}, and PeaTMOSS~\cite{jiang_peatmoss_2024}.

\begin{enumerate}[leftmargin=0.5cm]
    \item \textbf{HF Model Metadata} provides a snapshot of 10,406 HuggingFace model metadata as of 11/2022, including details of model label, README and length 
 of each README file~\cite{vanstrien_hf_model_metadata}.
    \item \textbf{PTMTorrent} encompasses a snapshot of five model hubs, totaling 15,913 PTM packages as of 08/2023, all formatted in a uniform data schema to facilitate cross-hub mining~\cite{jiang_ptmtorrent_2023}.
    \item \textbf{HFCommunity} is an offline relational database constructed from the data at the Hugging Face Hub. It allows for queries on the repositories hosted within the Hugging Face platform~\cite{ait2023hfcommunity}.
    \item \textbf{PeaTMOSS} offers comprehensive metadata on 281,638 PTM packages as of 10/2023, including 281,276 from Hugging Face and 362 from PyTorch Hub, along with details on 28,575 GitHub projects that use PTMs as dependencies and 44,337 links from these GitHub repositories back to the PTMs they depend on~\cite{jiang_peatmoss_2024}.
\end{enumerate}

In this work, we primarily used the PeaTMOSS dataset, as the accessibility of metadata allowed for easier measurements.
We also used the HF Model Metadata and PTMTorrent datasets for longitudinal trends in the turnover metric (\cref{sec:RQ2-Claim2}), and an April 2024 recent snapshot of the most popular models from the Hugging Face Hub API for the same reason.
The detailed data used for each measurement are presented in their respective sections.

\subsubsection{Traditional package datasets.}
\label{sec:Dataset-TradSW}
To directly compare measurements between the PTM registry and traditional Software Package Registries (SPRs), we utilized the Ecosyste.ms software package dataset for traditional packages, following the approach outlined in prior work~\cite{schorlemmer2024signing}. 
This dataset provides a set of free and open resources for those working to sustain and secure open source software. 
    \code{Ecosyste.ms} publishes open data and APIs that map software interdependencies, as well as providing data on the usage, creation, and potential impact of packages. 
In this work, we used the version of the dataset from October 2023, as this is the closest in time to when the PeaTMOSS dataset was created. This includes the usage of data from two software package registries: \pypi and NPM.

\subsection{$C_1$: Transformers increases the accessibility of PTM creation and downstream reuse.}
\label{sec:RQ2-Claim1}

{
\renewcommand{\arraystretch}{0.9}
\begin{table*}
    \centering
    \caption{
    This table displays the relationships between Qualitative Claims, the Metric(s) used to evaluate them, and the Hypothesized Results. 
    Where applicable, a reference is made to the traditional software prior work and metric that informed our metric. 
    Note that not every metric contains a reference to a traditional software prior work that uses this metric as no analog exists. 
    For those metrics, we utilize the Goal-Question-Metric process to develop metric(s) associated with the claim. 
    Since the goal of these measurements is to substantiate existing claims from prior work, we include a hypothesis about what the quantitative measurement will be if the claim is true.
}

\small
    \begin{tabular}{m{0.35\linewidth} m{0.28\linewidth} m{0.3\linewidth}}
    \toprule
    \textbf{Qualitative Claim} & \textbf{Metric} & \textbf{Hypothesized Result} 
    
    \\
    \midrule
    $C_1$: The Transformers library improves the process of PTM evolution. & Preservation rate of Libraries to descendents. & The preservation of Transformers as a library to its descendents will be greater than that of other libaries.\\

\addlinespace
    $C_2$: The traditional attribute of Popularity affects PTM selection and reuse more than Maintenance and Quality. & Turnover of top Packages over time \cite{dey_are_2018}. & Models with high popularity remain popular over time (``rich get richer''). \\
    \addlinespace
     & Descendent amount of models & Models with high popularity have a larger number of descendent models. \\
    \addlinespace
    $C_3$: Documentation quality impacts model selection. & Popularity of PTMs based on their documentation quality. & Models with more information are more discoverable and therefore more popular \cite{zerouali2018empirical}. \\
    \addlinespace
    \midrule
    \addlinespace
    $C_4$: DL-specific attributes such as model architecture, performance, reproducibility, and portability affect PTM selection and reuse. & Popularity of PTMs based on their Attributes \cite{zerouali2018empirical}. 
    
    & Popular architectures, high performance, description of where the PTM came from, ease of use, and size all impact model popularity. \\

\addlinespace
    \bottomrule
    \end{tabular}
    \label{tab:Claim-Metrics}
\end{table*}
}

\subsubsection{Methodology.}
We present the metric we developed for claim 1.
\textbf{Metric 1: Preservation rate of libraries to descendents:}
Some models support multiple libraries.
If the claim holds, then descendants of those models should make use of (and thus support) the more reuse-friendly libraries.
We examine each library $L$ in turn to assess how frequently its descendants continue to use it.
We consider:
  the library $L$ being assessed,
  the set of base models $B$ that use library $L$ and at least one other library,
  the set $D_b$ of direct descendants of base model $b \in B$,
  and
  a function $S(d, L)$ that returns 1 if descendant  $d \in D_b$  supports library $L$, otherwise 0.
We can then calculate the \emph{preservation rate}~$P_L$ of a library $L$ as: 

\begin{equation}
P_L = \frac{\sum_{b \in B} \sum_{d \in D_b} S(d, L)}{\sum_{b \in B} |D_b|}
\end{equation}

The numerator is the total descendants supporting library $L$, calculated as:
$\sum_{b \in B} \sum_{d \in D_b} S(d, L)$.
The denominator is the total count of direct descendants across all base models in set $B$: $\sum_{b \in B} |D_b|$.
This equation will give the percentage of direct descendants that support library $L$. 
The raw count of descendants supporting the library can be found using the numerator.

\subsubsection{Results.}
\begin{figure}[h]
    \centering
    \includegraphics[width=\columnwidth]{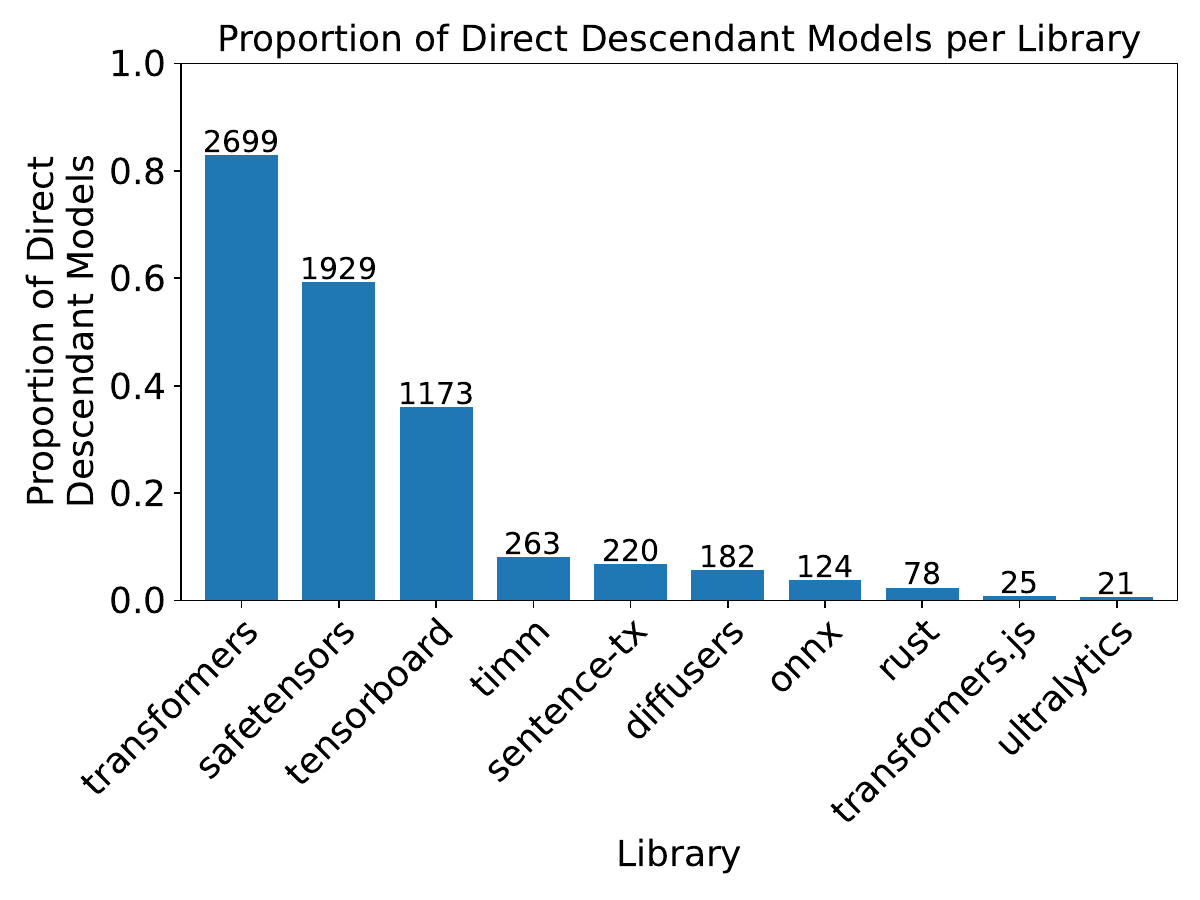}
    
    \caption{
    The usage proportion of the top-10 libraries that PTMs utilizing at least two different libraries use on \hf. 
    For PTM packages that leverage support at least two libraries, most packages support the transformers library, followed by the \hf promoted SafeTensors library. 
    Most other libraries have little usage in comparison.  
    In contrast to previous work~\cite{castano_analyzing_2023}, PyTorch is not one of the most popular library to be supported when a PTM package supports more than one library.
    }
    \label{fig:FORKS}
\end{figure}

\textbf{Metric 1: Preservation rate of libraries to descendents:}
As depicted in~\cref{fig:FORKS}, the Transformers library has the highest survival rate from a parent model to its descendant among all libraries used.
Over 80\% of descendant models continue to employ the Transformers library, establishing it as the preferred choice for generating descendant models.
Contrarily, PyTorch was not favored among PTM descendants, a departure from claims within our systematic literature review that both Transformers and PyTorch are dominant on Hugging Face~\cite{castano_analyzing_2023}.
Instead, our findings highlight the prevalence of the Transformers and SafeTensors libraries.
This suggests a community shift towards prioritizing security, as SafeTensors is designed to replace the commonly used Python pickle library with a more secure container for deploying models.
Our results suggest that PTM synthesizers are willing to compromise on functionality and portability to ensure the distribution of more secure PTM packages.
Building upon these observations, our findings reveal an opportunity for PTM platforms to further enhance security features.
This could involve implementing a visible security rating system for models and developing compatibility layers between secure libraries and those offering greater functionality or portability, encouraging developers to prioritize secure practices without sacrificing functionality or portability.

\subsection{$C_2$: Popularity Affects PTM Selection and Reuse More than Other Trad'l. Attributes}
\label{sec:RQ2-Claim2}

Our claim interpretation is that popular models are more likely to be used, so that the ``rich get richer''.
We examined this claim with two metrics.
First, we measure the stability (\ie non-turnover) of the top PTM packages over time, expecting it to be low (popular packages are used directly).
Second, we measure the correlation between popularity and the number of descendants of top PTM packages, expecting it to be positive (popular packages are fine-tuned).

\subsubsection{Methodology.} \label{sec:RQ2-Claim2-Methods}
We present the metrics developed for claim 2.

\textbf{Metric 2: Turnover of Top PTMs:}
Drawing on prior work characterizing the stability of top packages over time \cite{dey_are_2018}, we measured the \emph{top-$K$ turnover} for each registry.
Let $S_{current}$ be the set of $K$ most popular packages in the current snapshot, and $S_{last}$ be the set of top $K$ packages in the last snapshot.
We also consider $S_{history}$, the set of packages in any snapshot before $S_{last}$.
We then distinguish three categories for packages in $S_{current}$:

\begin{enumerate}[leftmargin=0.5cm]
    \item \textbf{Remained:} Packages that were in both $S_{last}$ and $S_{current}$.
    \begin{equation}
    \text{Remained} = S_{last} \cap S_{current}
    \end{equation}
    
    \item \textbf{Newcomers:} Packages in $S_{current}$ in no previous snapshot.
    \begin{equation}
    \text{Newcomers} = S_{current} \setminus (S_{history} \cup S_{last})
    \end{equation}

    \item \textbf{Returning:} Packages that were once in the top 1000 ($S_{history}$), were not in $S_{last}$, but are in $S_{current}$ again.
     
    \begin{equation}
    \text{Returning} = (S_{history} \setminus S_{last}) \cap S_{current}
    \end{equation}
\end{enumerate}

For the measurement, we defined popularity by the number of downloads, and examined the top-1000 packages.
We obtained snapshots across four dates for both traditional software and PTMs.
Data came from several HuggingFace snapshots (datasets: HF Model Metadata--June 2022, PTMTorrent--May 2023, PeaTMOSS--Oct. 2023). 

We took a current snapshot of the top 1000 PTMs directly from Hugging Face (April 2024).
We used the Ecosyste.ms dataset (NPM, PyPI) for a comparison with traditional software package registries.

\textbf{Metric 3: Number of Descendents of Top Packages:}
Our second measure of the impact of popularity on reuse was the number of descendent models.
In this case, we defined descendent models as a downstream model that is fine-tuned and references the original model as a base model.
We compare the number of descendent models with the popularity of model and determine the strength of correlation between them --- the claim implies it should be positive.

For the measurement, we again defined popularity by the number of downloads.
The descendent-base relation is available from the PeaTMOSS dataset, for the 15,000 most popular PTMs on Hugging Face.
Given that these models account for $\sim$99\% of the downloads in the snapshot, we believe this is representative.

We initially planned to compare our findings with traditional software, similar to Metric 2. 
However, identifying a direct counterpart to PTM descendants in traditional software proved challenging. 
We considered using GitHub forks and GitHub or registry dependencies as analogs, but each presented unique implications that complicated a direct comparison with PTM descendants.

\subsubsection{Results.}\label{sec:q2-results}

\begin{figure}[h]
    \centering
    \includegraphics[width=0.97\columnwidth]{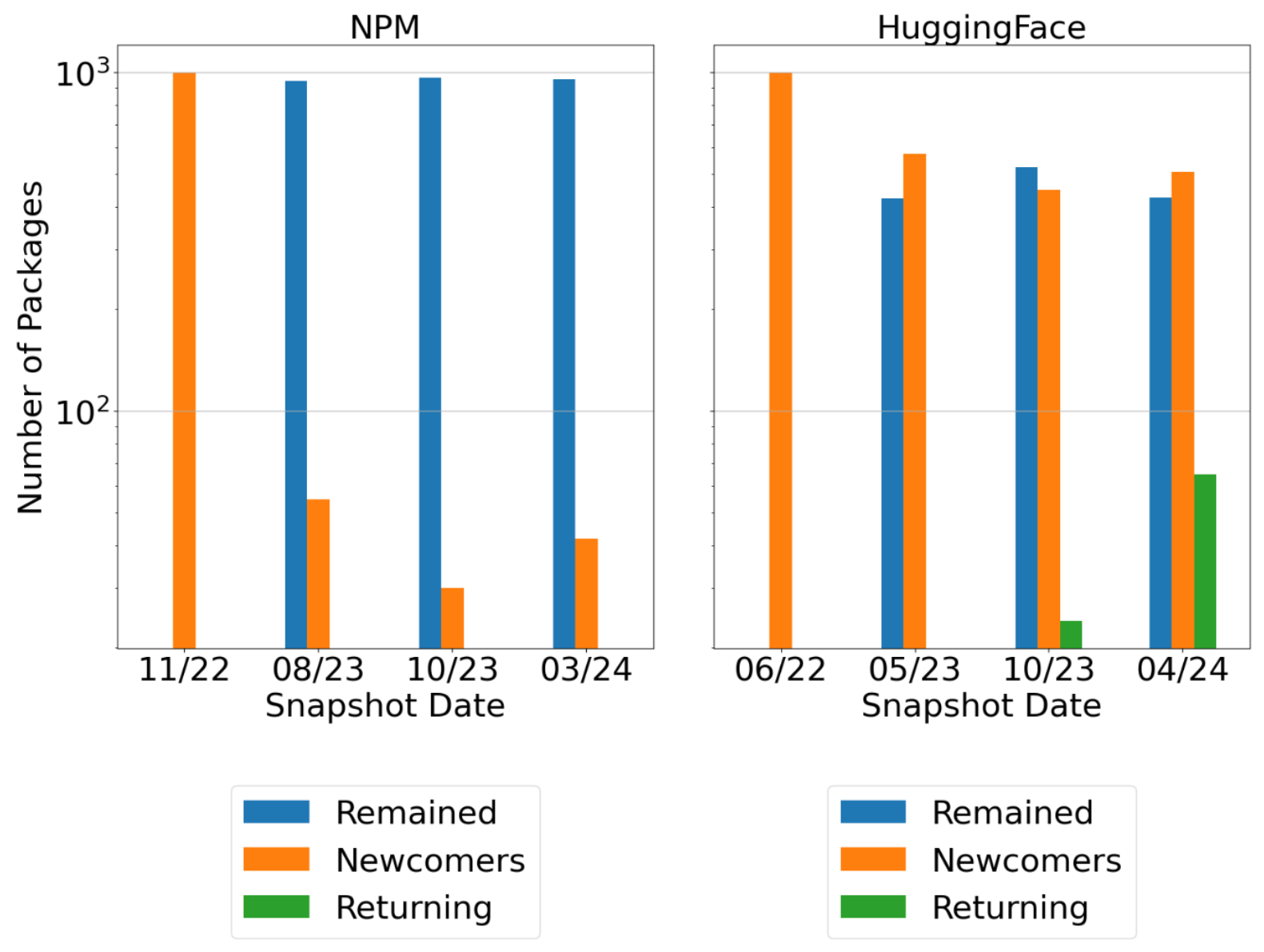}
    \caption{
    The turnover of the top-1000 Packages on Hugging Face, NPM. Note that in traditional software package registries such as NPM, the level of turnover is low. 
    Hugging Face has a much larger amount of packages re-enter the Top 1000 as shown by the larger green bars in the second and fourth snapshot compared to NPM, with few returners as shown by the relative lack of green bars. 
}
    \label{fig:turnover}
\end{figure}

\textbf{Metric 2: Turnover of Traditional Software and PTM registries:}
\label{sec:turnover}
\cref{fig:turnover} shows the results for Hugging Face and NPM for comparison.
The Hugging Face data does not match our interpretation of the claim, with about half of the top-1K Hugging Face PTMs turning over in each snapshot.
This high turnover rate suggests that packages on Hugging Face have a shorter lifespan, indicating a dynamic PTM environment where the requirements and preferred models frequently change~\cite{gopalakrishna_if_2022}.
In contrast, the traditional registries of PyPI and NPM showed stability with their most popular packages.
While 2535 distinct packages rotated through HuggingFace's top-1K, only 1127 featured for NPM.

The rapid turnover of top models on Hugging Face, in contrast to the stability seen in traditional software packages, reveals that PTM needs are constantly evolving, unlike the more established requirements in conventional software domains.
This dynamic environment sees newer models supplant older ones, driven by innovation and rapid adaptation to shifting demands.
Further investigation could analyze the lifecycle of these PTMs to determine whether they are newer models briefly appearing or established ones losing prominence, as well as identifying the traits of PTMs or maintainers who consistently stay within the top rankings.
These findings have significant implications for improving PTM reuse platforms.
Hugging Face and similar platforms should implement robust version control and model lineage tracking systems, enabling users to navigate the rapidly changing landscape of popular models.
Users need tools to easily track model evolution, understand version relationships, and identify relevant updates for their specific use cases.
Such enhancements would empower users to make informed decisions about model selection and upgrades, minimizing their technical lag in AI advancements.

\begin{figure}
    \centering
    \includegraphics[width=0.93\columnwidth]{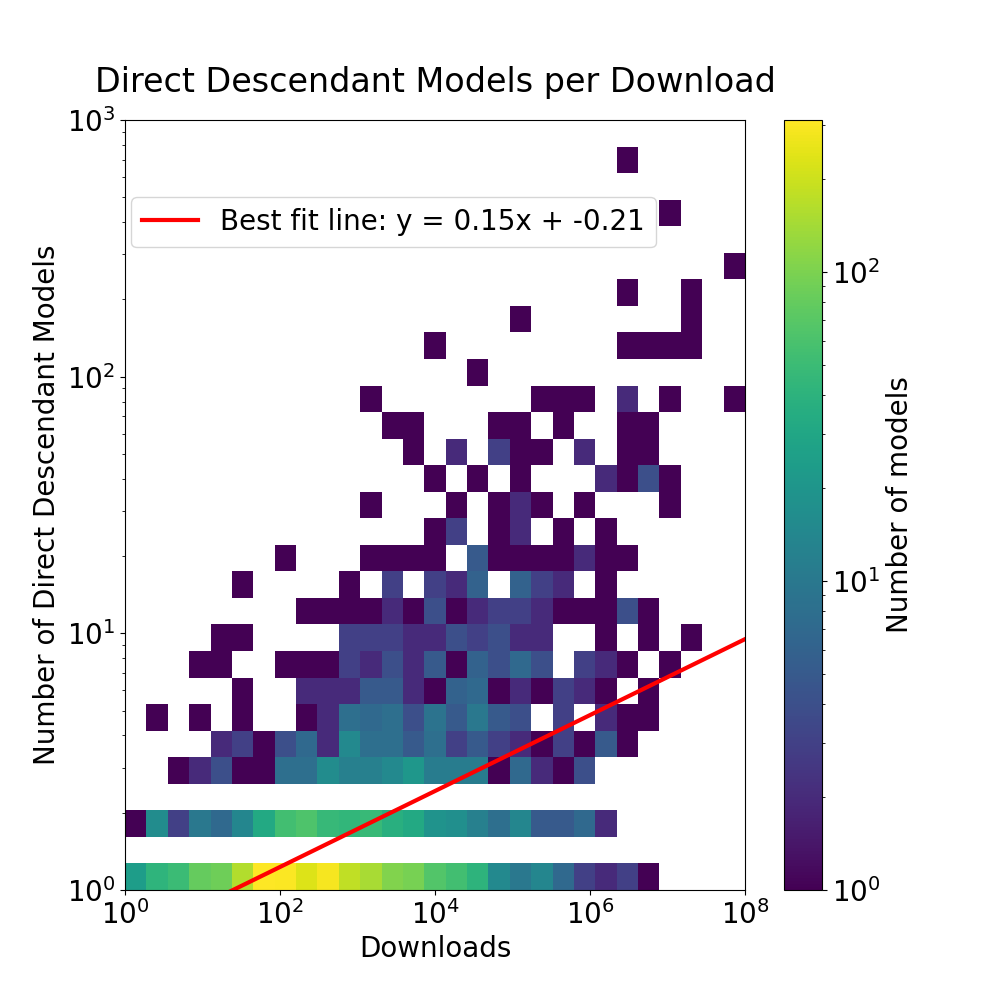}
    \caption{
    A heatmap showing the correlation between Downloads and Number of Direct Descendent Models. Note that this is a log-log plot. The red-line displays the best-fit relationship between downloads and descendant count, with a slope of $0.15$. This positive correlation suggests that models with higher download counts generally have a larger number of direct descendants, indicating that popular models tend to be reused more frequently in derivative work.
    }
    \label{fig:heatmap}
\end{figure}

\textbf{Metric 3: Number of Descendents of Top Packages:}
\label{sec:descendents}
\cref{fig:heatmap} shows the results for this metric.
We observe a weak positive correlation between popularity (downloads) and the number of descendent models.
While less significant than Metric 2's findings, our results highlight the need for specialized tools in PTM reuse and guide future research.
A more sophisticated recommendation system for PTM platforms is crucial, especially for identifying models suitable for further development.
Popularity alone is insufficient for engineers creating new models; more nuanced criteria are necessary.
Further analysis should consider factors like model architecture, application diversity, descendant generation patterns, and adaptation potential.
This approach could inform the development of informative recommendation systems that help users discover highly adaptable models across various domains and tasks, including those that might otherwise be overlooked due to lower popularity.

\subsection{$C_3$: Docs Quality Impacts Model Selection}
\label{sec:RQ2-Claim3}

Our interpretation of Claim 3 is that PTMs with better documentation will be more popular.

\subsubsection{Methodology.} \textbf{Metric 4: Documentation Quality:}
Prior work has examined documentation quality in many ways~\cite{gong_what_2023, tang2023evaluating}.
\JD{Help}
We used those ideas to develop our measure of quality.
The primary documentation for PTMs is called a ``model card'', which is similar to the README of a GitHub repository or the landing page of an NPM or PyPI package.
We considered two factors:
  (1) the completeness of the model card;
  and
  (2) the availability of metadata.

To measure completeness, we identified five typical sections found in highly popular PTMs such as Google's \textit{Bet base uncased}: \textit{Model Description}, \textit{Limitations}, \textit{How to Use}, \textit{Training}, and \textit{Evaluation}.
We scored model cards on an integer scale from 0 to 5 --- to receive a score of 5, a PTM's card needed all of these sections.
To check if each section was present, we queried OpenAI's ChatGPT-4 (Listing~\ref{lst:prompt4docquality}).

\begin{lstlisting}[caption={ChatGPT-4 prompt for evaluating model cards.}, label={lst:prompt4docquality}, breaklines=true, basicstyle=\ttfamily\footnotesize, frame=tb]
You will receive a model card and are expected to analyze it for the following details:
1. Model description: A description of the model itself
2. Limitations: Any limitations of the model
3. How to use: Instructions on how to use the model downstream
4. Training: Details of the training process or data
5. Evaluation: Reports on the model's performance evaluation

Please respond with a JSON object indicating whether each of these points is present with true/false.
Here is the model card to evaluate:
\end{lstlisting}

To measure the availability of metadata, we referenced the PeaTMOSS dataset, which extracted over 20 distinct pieces of metadata if present in a PTM's model card and associated configuration files.
We scored PTMs on an integer scale from 0 to the maximum, \ie the number of distinct pieces of metadata considered in the PeaTMOSS database schema. A perfect scoring PTM in this category would possess all available metadata according to PeatMOSS.

The PeaTMOSS dataset comprises extracted metadata from all models on Hugging Face and includes snapshots of the top 15,000 most popular models, each with over 50 monthly downloads. These models were further analyzed using an LLM, which extracted additional metadata from the model cards and the \code{config.json} files. Consequently, our analysis focused on these top 15,000 models to assess whether documentation quality influences model selection.

For this measure, we considered popularity in three ways:
  Downloads and Likes, according to Hugging Face,
  and
  Downstream Dependents, based on the mapping offered by the PeaTMOSS dataset.

To test our interpretation of the metric, we selected the top 1000 and bottom 1000 models from PeaTMOSS's set of 15,000 PTMs.
We evaluated their documentation quality, summing an overall documentation score as a (0,1) metric by normalizing the two components and weighting them equally.
Then we compared the distributions using a box-and-whisker plot and statistical tests.

\subsubsection{Results.}

\begin{figure}
    \centering
    \includegraphics[width=\columnwidth]{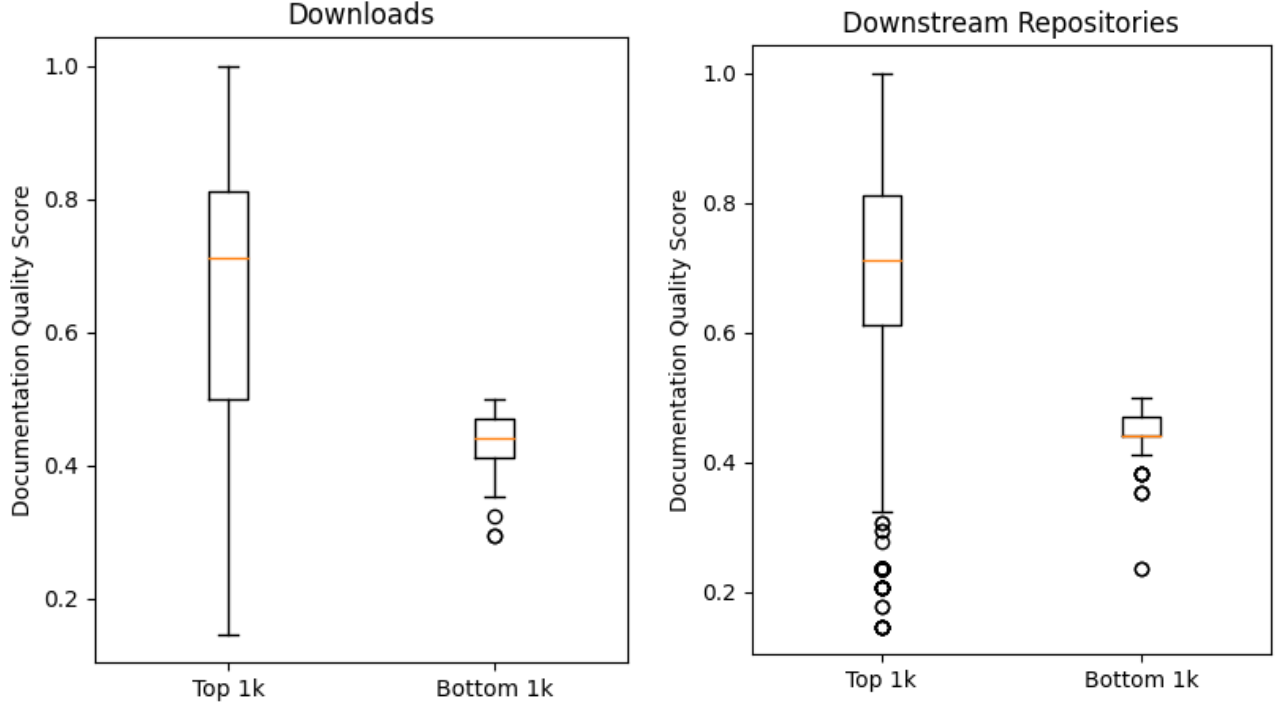}
    \caption{
    This figure shows the impact of documentation quality on model popularity using two popularity metrics: downloads and downstream reuse.
    The left box plot compares the documentation quality of the top 1,000 and bottom 1,000 models based on the number of downloads.
    The right box plot makes a similar comparison based on the number of downstream repositories.
    In both metrics, the top models demonstrate higher documentation quality scores than the bottom models, highlighting that models with better documentation are more popular and are reused more frequently.
    }
    \label{fig:DocQualityBoxWhisker}
\end{figure}

\textbf{Metric 4: Documentation Quality:} We evaluated the impact of documentation quality on the popularity of PTMs.
Two representative box and whisker plots are shown in~\cref{fig:DocQualityBoxWhisker}.
Specifically, the top 1000 most popular models consistently exhibited significantly better documentation than the bottom 1000 models ($p<0.01$.
This finding supports the claim that documentation quality significantly influences model selection.
If the relation is causative (documentation $\rightarrow$ popularity), then research can focus on developing tools and methods to enhance the documentation quality of models, supporting better usability and adoption.

\section{Threats to Validity}

We discuss three types of threats to validity~\cite{wohlin2012experimentation}, while considering the criticisms of Verdecchia \etal~\cite{verdecchia2023threats}.

\textbf{Construct Threats} are potential limitations of how we operationalized concepts.
In the systematic literature review (RQ1), we manually extracted claims from papers, which might introduce potential bias to our results.
As a mitigation, to improve objectivity two authors worked together on the process and the filtering of claims.
In the validation of non-quantified claims, we proposed metrics that seemed suitable based on our judgment.
To mitigate, we used multiple measures and leveraged existing metrics.

\textbf{Internal threats} are those that affect cause-effect relationships.
We emphasize that our approach for RQ2 is of the form:
  ``\textit{If claim $C$ is true, then measurement $M$ should show us that...}''.
In each case the measurement produced the expected result.
However, this result is correlative, not causative --- there may be a latent variable in each case, or the causative relationship may be reversed.

For example, in~\cref{fig:DocQualityBoxWhisker} we found that better documentation correlates with greater popularity.
It may be that the latent variable is performance, such that models become popular because they have good performance, and they accrue documentation because they are popular. 
When qualitative and quantitative claims agree, as is the case in this study, we learn both ``\textit{Why?}'' and ``\textit{How much?}''.

\textbf{External threats} may impact generalizability.
We recognize both immediate and longitudinal threats in this regard.
Immediately, we were interested in studying PTM reuse, but we only examined one registry for PTMs: the Hugging Face platform.
While this is by far the most popular and feature-rich platform, other platforms exist, such as PyTorch Hub (less popular), PapersWithCode (fewer features), and GitHub (not PTM-specific).
In terms of longitudinally, keep in mind that Hugging Face is itself relatively young --- created in 2016 and only seeing major use beginning in 2020 --- so developer practices may not have stabilized.
Lastly, we consider that the technologies and platforms that support PTMs are rapidly evolving, so current claims (whether qualitative or quantitative) may change over time and require ongoing reassessment.
As indicated in the discussion, this property suggests an opportunity for further research, but it also means that our findings may be unstable.

\section{Future Work}
\label{sec:discussion}

Our study provides the first systematic literature review of current knowledge about PTM reuse.
Another opportunity is a systematic \textit{comparison} of the software package registries associated with traditional software packages, and the software package registries associated with PTMs.
Jiang \etal observed similarities and differences in the reuse processes~\cite{jiang_empirical_2023} --- how and to what extent can we measure the differences?
As discussed in~\cref{sec:RQ2-Claim2-Methods} with respect to an analogue for PTM descendants, finding the limits of comparison (\eg appropriate measurements) between traditional vs. PTM software package registries is an open challenge.

In \cref{sec:q2-results}, we identified several potential improvements for PTM platforms, particularly in the areas of version control, model lineage tracking, and recommendation systems.
However, the specific attributes that these platforms should measure and analyze remain underexplored.
Our findings suggest that popularity is a factor in model selection and reuse, but not the sole determinant (cf. claim $C_4$).
Therefore, we recommend further research to uncover and quantify the deep learning-specific attributes that influence a model's reusability and adaptability.
Such research would provide valuable insights for developing more sophisticated tools and metrics for PTM platforms, enabling users to make more informed decisions when selecting models for reuse or further development.



The rapid development of PTM technologies presents an ongoing challenge for empirical research on PTM reuse.
The Hugging Face platform continues to grow exponentially~\cite{castano_analyzing_2023},
   the state of the art performance of models at all sizes continues to advance~\cite{paperswithcode}, 
   and the tooling available to adapt and deploy these models continues to improve~\cite{Jajal2024Interoperability}. \GKT{make sure to get the citations in.}
Empirical software engineering researchers need to stay abreast of the activity and volume of data in the context of PTM reuse. 
Given Hugging Face's dynamic growth, even data that is a few months old may not accurately reflect the current state of the platform, suggesting a need for regular snapshotting, which imposes significant storage requirements (beyond the already-substantial requirements of $>$50 TB). 
This highlights the need for tools that can provide real-time, incrementally-updated data to keep pace with rapid changes, ensuring that analyses remain relevant and reflective of the present situation in PTM reuse.

\section{Conclusion}



Pre-Trained Models are driving the next generation of software engineering.
We must understand (and optimize) engineers' PTM reuse practices.
This systematic literature review has synthesized existing knowledge about PTM reuse in the Hugging Face registry.
We also quantified certain claims, revealing unique dynamics within the PTM landscape compared to traditional software package registries.
These findings motivate further research to understand and optimize the distinctive aspects of PTM registries for PTM reuse, to meet the evolving needs of the software engineering community.

\section{Data Availability}
\label{artifact}

An artifact is available~\cite{github-artifact}, containing
 results of the SLR (RQ1)
 and
 code and results for claim quantification (RQ2).
Source: \url{https://github.com/PurdueDualityLab/ptm-quantify-esem-2024/}.

\section*{Acknowledgments}

Work was supported by NSF awards \#2107020 and \#2107230.

\clearpage
\balance

\bibliographystyle{ACM-Reference-Format}
\bibliography{esem}

\appendix

\end{document}